# Johnson(-like)-Noise-Kirchhoff-Loop Based Secure Classical Communicator Characteristics, for Ranges of Two to Two Thousand Kilometers, via Model-Line[1]

Robert Mingesz [(x)], Zoltan Gingl [(x)] and Laszlo B. Kish [(+)]

[(x)] *Department of Experimental Physics, University of Szeged, Dom ter 9, Szeged, H-6720 Hungary*

[(+)] *Department of Electrical and Computer Engineering, Texas A&M University, College Station, TX 77843-3128, USA*

**Abstract.** A pair of Kirchhoff-Loop-Johnson(-like)-Noise communicators, which is able to work over variable ranges, was designed and built. Tests have been carried out on a model-line performance characteristics were obtained for ranges beyond the ranges of any known direct quantum communication channel and they indicate unrivalled signal fidelity and security performance of the exchanged raw key bits. This simple device has single-wire secure key generation and sharing rates of 0.1, 1, 10, and 100 bit/second for corresponding copper wire diameters/ranges of 21 mm / 2000 km, 7 mm / 200 km, 2.3 mm / 20 km, and 0.7 mm / 2 km, respectively and it performs with 0.02% raw-bit error rate (99.98 % fidelity). The raw-bit security of this practical system significantly outperforms raw-bit quantum security. Current injection breaking tests show zero bit eavesdropping ability without triggering the alarm signal, therefore no multiple measurements are needed to build an error statistics to detect the eavesdropping as in quantum communication. Wire resistance based breaking tests of *Bergou-Scheuer-Yariv* type give an upper limit of eavesdropped raw bit ratio of 0.19 % and this limit is inversely proportional to the sixth power of cable diameter. *Hao*'s breaking method yields zero (below measurement resolution) eavesdropping information.

## Introduction.

Recently, a totally secure classical communication scheme, a statistical-physical competitor of quantum communicators, was introduced, see Figure 1, [1,2], the Kirchhoff-loop-Johnson(-like)-Noise (KLJN) communicator, with two identical pairs of resistors and corresponding Johnson(-like) noise voltage generators. The KLJN communicators are utilizing the physical properties of an idealized Kirchhoff-loop and the statistical physical properties thermal noise (Johnson noise). The resistors (low bit = small resistor, high bit = large resistor) and their thermal-noise-like voltage generators

[1] The results were featured in the technology section of the New Scientist magazine by Jason Palmer, "Noise keeps spooks out of the loop", May 23, 2007. http://www.newscientisttech.com/article.ns?Id=mg19426055.300

(thermal noise voltage enhanced by a pre-agreed factor) are randomly chosen and connected at each clock period at the two sides of the wire channel. A secure bit exchange takes place when the bit states at the two line ends are different, which is indicated by an intermediate level of the *rms* noise voltage on the line, or that of the *rms* current noise in the wire. The most attractive properties of the KLJN cipher are related to its security [1,3] and to the extraordinary robustness of classical information when compared that to the fragility of quantum information. To provide security against arbitrary types of attacks, the instantaneous currents and voltages are measured at both end by Alice and Bob and they are published and compared. In the idealized scheme of the KLJN cipher, the passively observing eavesdropper can extract zero bit (zero-bit security) of information and the actively eavesdropping observer can extract at most one bit before getting discovered (one-bit security) [1]. The system has a natural zero-bit security against the *man-in-the-middle attack* which is a unique property among secure communicators [3]. The KLJN system has recently became network-ready [4]. This new property [4] opens a large scale of practical applications because the KLJN cipher can be installed as a computer card [4], similarly to Eternet network cards. Other practical advantages compared quantum communicators are the high speed, resistance against dust, vibrations, temperature gradients, and the low price [1]. It has recently been shown [5] that the KJLN communicator may use currently used wire lines, such as power lines, phone lines, internet wire lines by utilizing proper filtering methods.

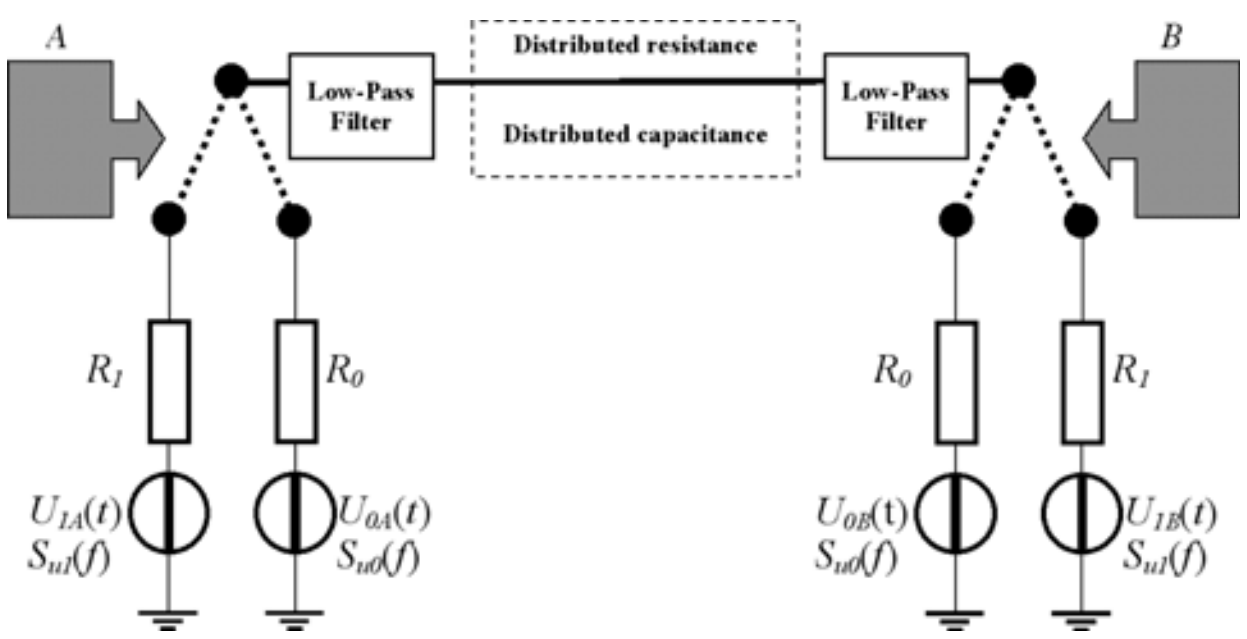

**Figure 1.** The outline of the KLJN communicator. The instantaneous current and voltage data are measured and compared at the two ends and, in the case of deviance; the eavesdropping alarm goes on and the currently exchanged bit is not used. The low-pass line filters are necessary to protect against out-of-alarm-frequency-band breaking attempts and false alarms due to parasite transients. False alarms would occur due to any wave effect (transient or propagation effects), illegal frequency components or external disturbance of the current-voltage-balance in the wire.

Concerning the security of the KLJN system, the basic rule of any physical secure communication holds here, too: *If you compromise it you lose it*. Though there have been several attempts to break in the KJLN line, so far, no proposed method has been able to challenge the total security of the idealized KLJN system. All these breaking attempts have used certain assumptions which are directly violating the basic model of the idealized KLJN method. These assumptions are as follows:

*i.* Allowing non-negligible wire resistance (Bergou [2], Scheuer and Yariv [6]). For the evaluation of this claim at practical conditions, see the response in [7].

*ii.* Allowing high-enough bandwidth for wave (transient propagation) effects (Scheuer and Yariv [6]). This possibility has been explicitly excluded since the original article [1]. For further refusal of this claim, see the response in [7]. Moreover, *and this is the most direct practical argument*, in this case the standard current and voltage protection of the KLJN system would *immediately alarm and shut down the communication* because the existence of transient and wave effects *per-definition* yield different instantaneous current and/or voltage values at the two ends.

*iii.* Assuming inaccuracy of (noise) temperatures (Hao [8]). For the theoretical refusal of this claim, see [9].

*iv.* Utilizing practical inaccuracy of resistors (Kish [9]). For the mathematical evaluation of this claim at practical conditions, see [9].

Let us here consider an analogy in the field of quantum communication. Though all the practical quantum communicators suffer from the inability of producing *totally single-photon* output and this fact is unavoidably compromising the security of practical quantum communicators, we do not say that the idealized/mathematical quantum communicator schemes are insecure. In discussing the unconditional security of idealized quantum communication schemes, we suppose that single-photon sources do exist, which is a similar though more difficult claim than to suppose the existence of zero wire resistance (c.f. superconductors), and then we conclude that the quantum system is therefore unconditionally secure, at least conceptually. Therefore the practical claim that idealized single-photon source does not exists cannot compromise the claim about the security of an idealized, conceptual quantum communicator. On the other hand, as soon as the security of a *practical communicator system* is discussed, the aspects non-ideal single-photon sources, optical fiber absorption, etc, for quantum communicators and similarly the aspects (*i-iv*) listed above for the KLJN systems cannot be neglected any more and the practical design must keep these effects under control.

In the present paper, we report the first experimental realization and tests of the KLJN secure communicator system. We test and analyze the most important breaking methods of the ones listed above. The tests indicate an unrivalled *beyond-quantum security level* of the realized system.

## 2. The realized secure classical physical communicator

The generic schematic of practical KLJN secure classical communicators is shown in Figure 2. The logic-low (L) and logic-high (H) resistors, $R_0$ and $R_1$, respectively, are randomly selected beginning of each clock period and are driven by corresponding Johnson-like noise voltages. The actual realizations contain more filters and amplitude control units (not shown here). The thick arrows mark computer control and data exchange. The controlling computer (not shown) has a regular network connection with the computer of the other KLJN communicator (not shown).

The realized pair of KLJN communicators [1-4] was designed and built for variable ranges. The circuit realization used Digital Signal Processor (DSP) and analog technology, see the outline in Figures 2 and 3. The details not described here are part of intellectual property which we cannot disclose at this stage. The computer control parts of the communicator pair have been realized by ADSP-2181 type Digital Signal Processors (DSP) (Analog Devices). The digital and analog units were placed on two separate computer cards for easier variability during tuning up. The communication line current and voltage data were measured by (Analog Devices) AD-7865 type AD converters with 14 bits resolution from which 12 bits were used. The DA converters were (Analog Devices) AD-7836 type with 14 bits resolution. The Johnson-like noise was digitally generated in the Gaussian Noise Generator unit where digital and analog filters truncated the bandwidth in order to satisfy the KLJN preconditions of removing any spurious frequency components. The major bandwidth setting is provided by an 8-th order Butterworth filter with sampling frequency of 50 kHz. The remaining small digital quantization noise components are removed by analog filters.

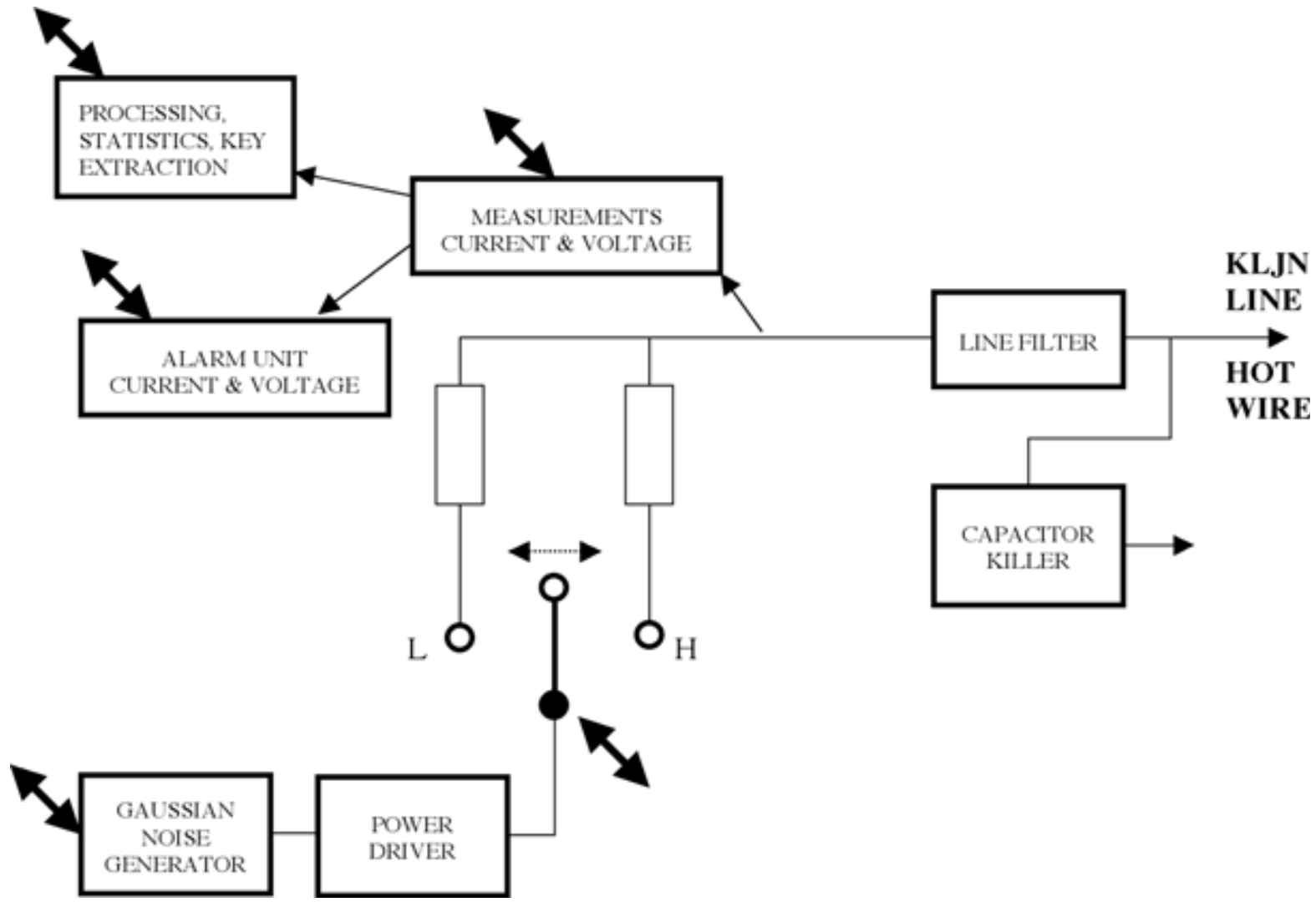

**Figure 2.** Generic schematics of practical KLJN secure classical communicators. The low (L) and high (H) resistors are randomly selected beginning of each clock period and are driven by corresponding Johnson-like noise voltages. The actual realizations contain more filters and amplitude control units (not shown here). The thick arrows mark computer control and data exchange. The controlling computer (not shown) has a regular network connection with the computer of the other KLJN communicator (not shown). Regarding the communication, a bare-wire line within the given ranges, in the frequency band relevant to the given range and with the driving resistor values described in the text, can be represented by a resistor and a capacitor and, below 1% of the ranges, as a simple resistor. With capacitor killer and twisted pair (two coaxial cables) arrangement, it can be represented by a single resistor throughout the earth globe in the relevant frequency bands.

The experiments were carried out on a model-line, with assumed cable velocity of light of $2*10^8$ m/s and capacitor killer arrangement, with ranges up to 2000 km (which is far beyond the range of direct quantum channels, or of any other direct communication method via optical fibers). Under these conditions, and with the applicable KLJN bandwidths and copper wire diameters given below, the line model is a single resistor with about 200 Ohm resistance. The corresponding copper wire diameters are reasonable practical values for the different ranges are 21 mm (2000 km), 7 mm (200 km), 2.3 mm (20 km) and 0.7 mm (2 km). Beyond 2000 km, the present realization would require non-practical wire diameters. Inductance effects are negligible with the selected resistance values, $R_0$ and $R_1$, at the given ranges and the corresponding bandwidths. A bare-wire line within the given communication ranges; in the frequency band relevant to the given range and with the driving resistor values described in the text, can be represented by a resistor and a capacitor and, below 1% of the ranges, as a simple resistor. With capacitor killer and twisted pair (two coaxial cables) arrangement, it can be represented by a single resistor throughout the earth globe in the relevant frequency bands. The capacitor killer (Figure 2) is a well-know electronic solution for capacitance compensation to remove capacitive cut-off effects in coaxial cables by feeding the coat of a coaxial cable by a follower driven by the hot wire input of the coaxial cable.

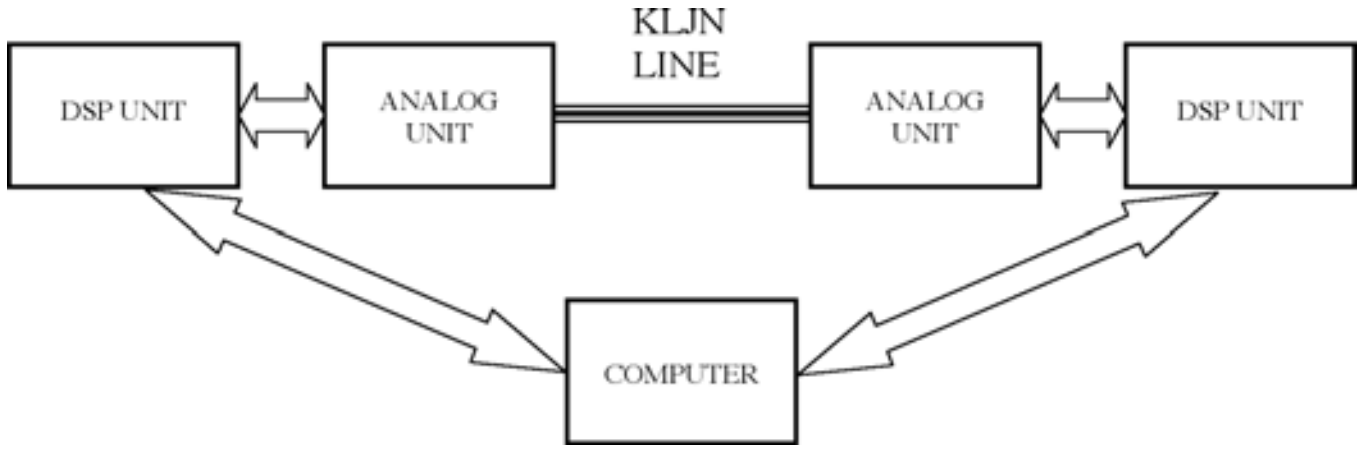

**Figure 3.** The schematics and arrangement of the realized and tested KLJN communicator pair. The KLJN line is a model line with capacitance compensation up to 2000 km range. Alternatively, freestanding wires can be used up to 1% of the nominal ranges with the given data. Alternatively, within the earth globe limits, arbitrarily longer distances can be reached with freestanding wires capacitor killer if lower driving resistances (and thicker cables) are used or the communication speed (bandwidth) is reduced.

The voltage noises are band-limited Johnson-like noise with power density spectrum

$$S_u(f) = KR \qquad (1)$$

within the noise bandwidth and zero elsewhere. The value of $K$ depends on the noise bandwidth determined by the actual range and it is selected so that the noise voltage of the greater resistor is 1 Volt for all noise bandwidths. This resulted in $S_u(f)$ values of the greater resistor 0.2, 0.02, 0.002, 0.0002 $V^2/Hz$ for ranges 2000, 200, 20 and 2 km, respectively. Wave effects in the stationary mode and information leak due to them are avoided by the proper selection of the noise bandwidth. The noise bandwidth is selected so that the highest possible Fourier component in the line is at frequency 10 times lower than the lowest frequency standing-wave mode in the line. That condition results in noise bandwidths 5, 50, 500 and 5000 Hz for ranges 2000, 200, 20 and 2 km, respectively. This condition and the statistical sample size within the clock period determine the speed (bandwidth) of communication. The sampling rate satisfies the Shannon limit thus the sampling frequency is two times greater than the noise bandwidth. The sample size within the clock period is equivalent to about statistically independent 50 data points (though the digital filtering is using about an order of magnitude greater actual sampling rate) and that results in a 50 times lower secure bit rate than the noise bandwidth because only 50% of the communicated bits are secure. Thus the device has bit rates of 0.1, 1, 10, and 100 bit/second for ranges 2000, 200, 20 and 2 km, respectively.

The resistors pairs at Alice and Bob have been selected so that, with the practical wire diameters given above, the wire resistance is about 2% of the loop resistance during secure bit exchange. The resistors pairs are $R_0 = 2$ kOhm and $R_1 = 11$ kOhm. The 11 kOhm resistor is composed by a serial 10 kOhm resistor and a 1 kOhm resistor with a 1 nF capacitor shunting their joint point to the ground to remove possible digital quantization noise. The 2 kOhm resistors are two serial 1 kOhm resistors with also a 1 nF capacitor shunting their joint point to the ground to remove possible digital quantization noise. These shunt capacitors were the first stages of the line filter. Because the security protection based on current and voltage comparison was effective up to 50 kHz bandwidth, the shunt capacitors described above and additional 1 nF capacitors at the two ends of the model line were satisfactory line filters in the model line. Note, actual practical arrangements with real wires automatically provide much larger capacitances. Furthermore, these capacitors would have removed possible switching spikes originating from capacitive coupling in the analog switches due to possibly unbalanced parasitic capacitors; therefore there were no detectable switching transients in the line. The 1 kOhm resistor at the generator driven end was also used as a probe to measure the current in the line.

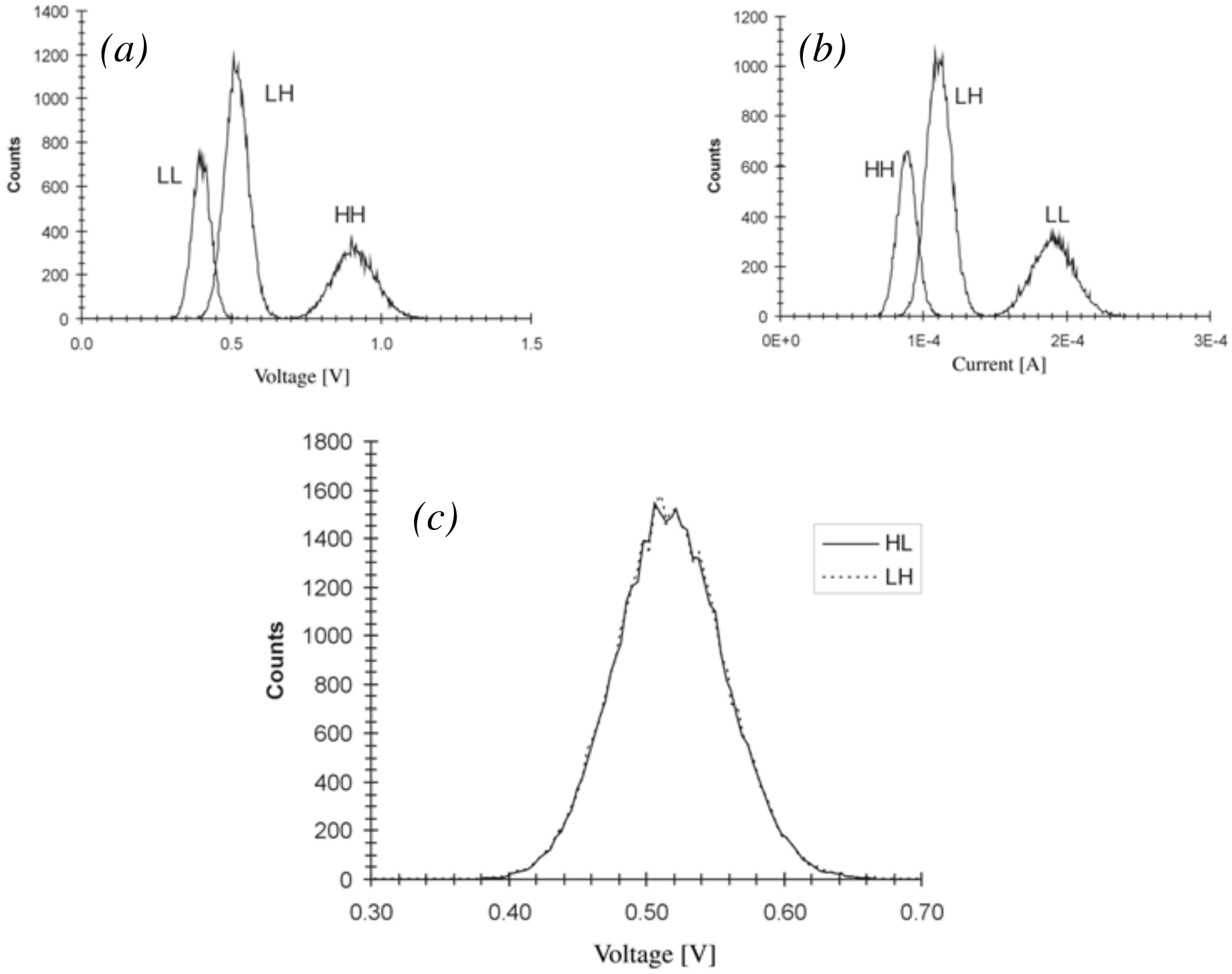

**Figure 4.** Empirical amplitude density functions (histograms) during the whole span of security checks utilizing 74497 clock cycles: (a) and (b) Voltage and current counts seen by Alice and Bob; (c) Voltage counts seen by Eve at one end of the line during the two different secure bit arrangements (*LH* and *HL*). These functions correspond to the situation when the bit arrangement is fixed (*LH* or *HL*) and then the two distribution functions are the voltage counts measured at the two ends of the line to execute a BSchY type of attack. It is obvious from the strong overlap of the two curves that Eve has virtually zero information even with fixed bit arrangement for 74497 clock cycles.

Transient wave effects at the end of clock period are avoided in the Gaussian Noise Generator unit by driving the envelope of the time functions of noise voltage and current to zero before the switching using a linear ramp amplitude modulation (via 8% of the clock duration); and the reverse process is done at the beginning of the next clock cycle after the switching of resistors. The switching of resistors takes place within a zero-current-zero-voltage window with duration of 9% of the clock period, and 0.8% before the start of the next clock cycle. Moreover a short pause (8 % of the clock time) with no data collection, except for security check, after the initial linear ramp at the beginning of stationary noise, is applied in order to avoid possible other types of transient/filter effects of stochastic nature. All these are done before the filtering process to avoid any spurious frequency components due to the linear ramp.

## 3. Experimental study of the security and the fidelity of the realized communicators

During the communication and security tests, the exchange of secure bits took place between Alice and Bob with 0.02% error rate, which is 99.98% fidelity, much beyond that of quantum channels. Figure 4 shows an illustrative example about the actual distribution functions and their statistics made over the over the total span of security checks (about 74497 clock cycles) and Figure 5 shows an example of the actual counts (working statistics) during a single clock cycle.

According to the security tests and the related evaluations the raw-bit security of this practical system has been successfully designed to outperform raw-bit quantum security while it still remains reasonably inexpensive. For the summary of security test evaluation, see Table 1.

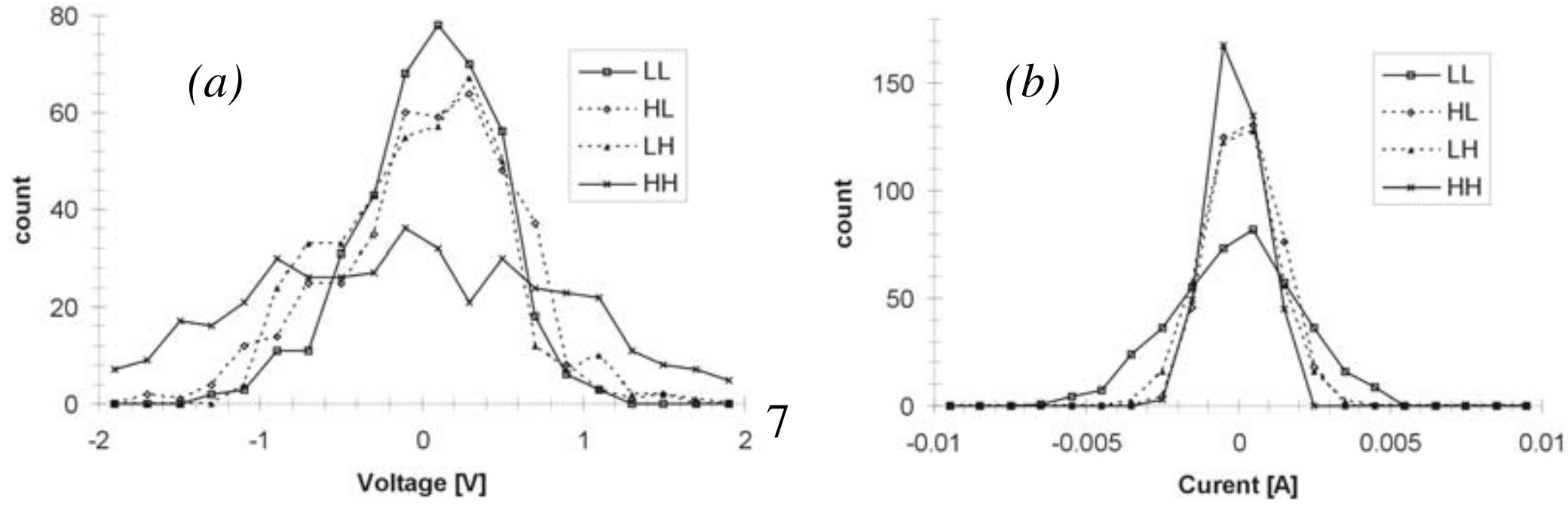

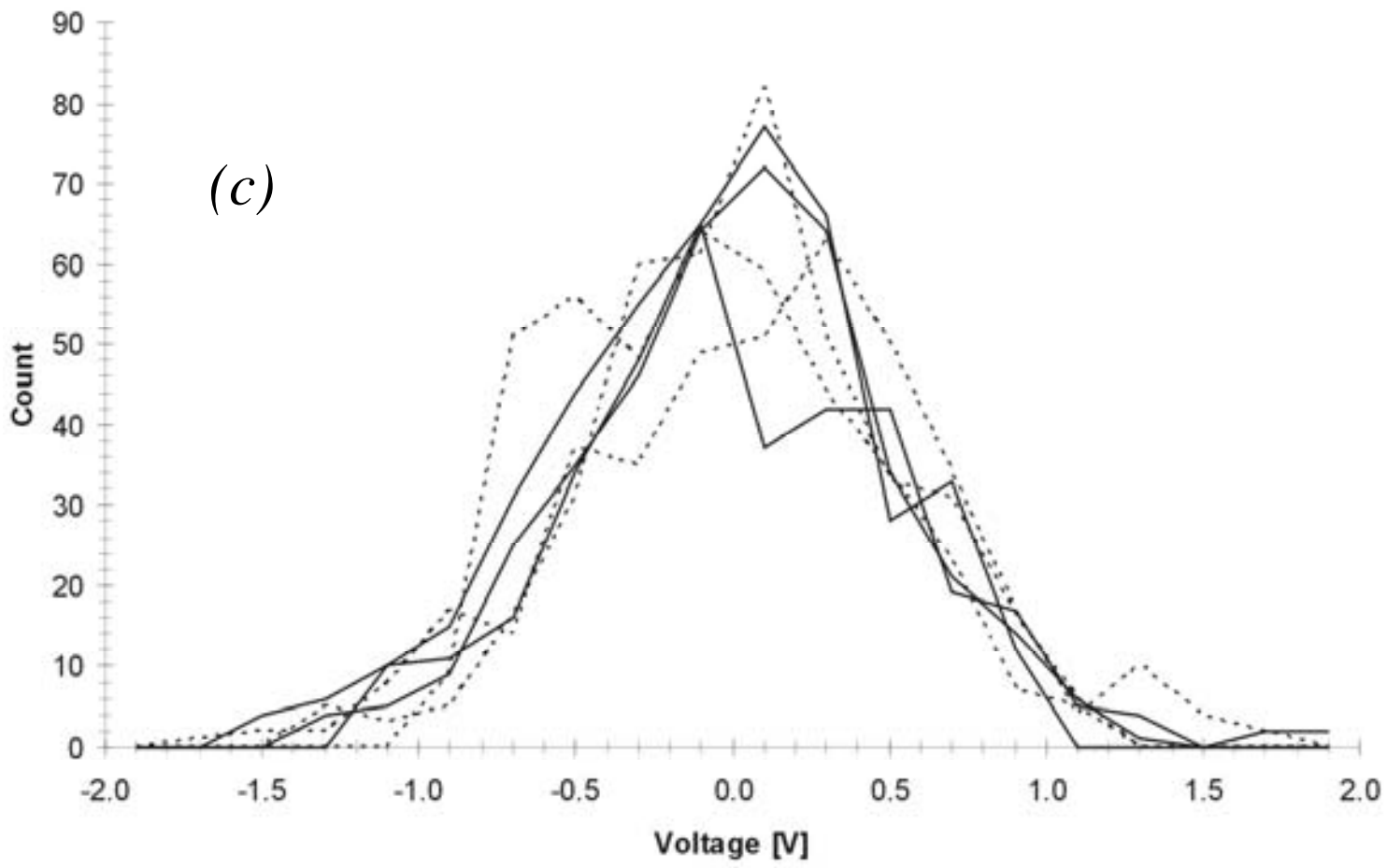

**Figure 5.** Empirical amplitude density functions (histograms) of the voltage (left window) and current (right window) during a single clock cycle: (a) and (b) Voltage and current counts seen by Alice and Bob; (c) Voltage counts seen by Eve at end of the line at the two different secure bit arrangements, *LH* and *HL*, three of each one. These functions correspond to the situation when the bit arrangement is fixed (*LH* or *HL*) and then the two distribution functions are the voltage counts measured at the two ends of the line to execute a BSchY type of attack. The poor statistics seen in figures (a) and (b) are enough for Alice and Bob to identify secure bit alignment with 0.02% error rate (99.98% fidelity). However when Eve tries to identify the bits from the two histogram recorded at the two ends of the line (see figure (c)) she must work with these distributions which are very stochastic, almost identical and totally overlapping with a less than 1% shift of their centers [7] which results in 0.19% eavesdropped bit / transmitted secure bit.

Breaking tests *(i)* of Bergou-Scheuer-Yariv (BSchY) type utilizing wire resistance [2, 6] were carried out by measuring and comparing the *rms* voltages at the two ends of the line and, during secure bit transfer, supposing *LH* or *HL* bit arrangement, accordingly. We have been using Shannon's channel coding theorem to evaluate the information leak and fidelity [7]:

$$\frac{C_{eav}}{C_{trans}} = 1 + p\log_2 p + (1-p)\log_2(1-p) \qquad (2)$$

where $C_{eav}$ is the upper limit of eavesdropped bits (supposing optimal coding/decoding), $C_{trans}$ is the total number of transferred secure bits, $p$ is the probability of correct assumption and (1-$p$) is the probability of erroneous assumption.

| | | |
|---|---|---|

| TYPE OF BREAKING | MEASURED NUMBER, OR RATIO, OF EAVESDROPPABLE BITS WITHOUT TRIGGERING THE CURRENT-VOLTAGE ALARM (TESTED THROUGH 74497 BITS) | REMARKS |
|---|---|---|
| BSchY (*i*) [2,6] attack in the present KLJN system | 0.19% | 0.00000019% at 10 times thicker wire (theoretical extrapolation). Arbitrarily can be enhanced by privacy amplification [12,13]; the price is reduced speed. |
| Hao (*iii*) [8] attack in the present KLJN system | Zero bit | Below the statistical inaccuracy. Considering the 12 bit effective resolution of noise generation accuracy, it is theoretically: < 0.000000006% |
| Kish (*iv*) [9] attack utilizing resistor inaccuracies in the present KLJN system | Zero bit | Below statistical inaccuracy. Theoretically, when pessimistically supposing 1% resistance inaccuracy, it is: < 0.01% |
| Current pulse injection (Kish) [1] in the present KLJN system | Zero bit | One bit can be extracted while the alarm goes on thus the bit cannot be used. |
| Breaking into a quantum channel by capturing and multiplying (noisy cloning) photons and restoring them in the line. | 1000 - 10000 bits | Because only the change in the detection error rate is able to uncover Eve. And that needs to build very good statistics. |
| Breaking into a quantum channel by randomly stealing photons, multiplying (noisy cloning) and feeding back to the line. | >1% | Arbitrarily can be enhanced by privacy amplification [12,13]; the price is reduced speed. |

**Table 1.** Summary of the security analysis of the realized KLJN communicator pair and a quick comparison with quantum security.

The experimental test of this practical information leak through 74497 clock cycles shows $p = 0.526$ which yields $C_{eav} / C_{trans} = 0.19\%$ and that is less than practical quantum bit-leak when photons are randomly stolen, copied (noisy cloned) and replaced; a method by which Eve can easily extract 1% of the bits while staying hidden at practical situations. However to reach this 0.19% information leak Eve must know/use the optimal decoding of data, in accordance with Shannon's channel coding theorem, and that is usually unknown. Moreover, the location of correctly eavesdropped bits is random and that is also a significant deficiency as compared to the random stealing of photons in quantum channels where the location of the correctly eavesdropped bit is known (with reasonable accuracy). A simple improvement possibility is served by the fact that the BSchY bit-leak scales with the reciprocal of the sixth power of cable diameter [10] and that means that doubling cable diameter would reduce this 0.19% leak by a factor of 64 and a diameter increase of a factor of 10 (possible only for 20 km or shorter ranges due to practical limits of cable diameter) would yield a million times reduction of the bit leak. Note, using additional software tools can provide arbitrarily strong reduction of the bit-leak; this is the *privacy amplification* technique [12,13] developed for quantum communication by Bennett and coworkers. However, the price of that is reduced speed.

Hao's *(iii)* breaking method yields zero (non-measurable) information for the eavesdropper due to the 12 bits effective accuracy of current and voltage measurement and the limited statistics during the clock period. Considering the 12 bit effective resolution of noise generation accuracy, theoretical estimation based on Eq. (2), see [7,9], yields an information leak of less than 0.000000006%.

Kish *(iv)* [9] attack utilizing resistor inaccuracies in the present KLJN system yielded zero extractable information because of the accurate choice of resistors at the two line ends. Supposing 1% resistance inaccuracy, the bit leak is less then 0.01%, according to estimations based on Eq. (2), see [7,9].

Current pulse injection breaking attempts (Kish) [1] were performed by discharging a capacitor (1 nF, 5 Volt) on the line to inject a short current pulse (relevant for the line filter's high-frequency cutoff) and to use the current distribution to read out the resistance values. These breaking attempts are able to extract just a single bit of information while the current-alarm goes on. Therefore, as expected, this type of break yields zero useful bit extraction. The smallest detectable current difference at the two ends of the line current was 0.025% of the *rms* current in the line. Note, less sensitivity would have produced the same practical security level however, due to the given resolution of the AD converters, we evaluated the performance with this resolution.

Finally, we would like to note that, because breaking tests by the *man-in-the-middle attack* require involved hardware and significantly more efforts, they will be performed later.

## 4. Comments on the speed of communication

So far, trusted experimental data published on the speed of quantum communicators is very similar to that of the practical KLJN speed described above, or sometimes closer to its theoretical single-line limit which is about 10 times faster. It is important to note that, in the literature, there are various claims about the speed of quantum communication however one must take seriously only those experimental works, which are cautious about security, and keep/use only the really secure bits. This process results in a significant bandwidth limitation. Zeilinger and coworkers have recently reported state-of-the-art quantum communication with enhanced security, which requires entangled photon pairs, via the range of 144 km [15]. During a run of 75 second, they obtained 789 raw bits (coincidences) [15]. This is a speed of about 10 bit/second however their information leak with these bits is 4.8% and that is 25 times higher than that of the KLJN communicator (0.19%) at the given wire resistance. After using their privacy amplification algorithm on these bits, they achieved a secure key (final information leak data are not provided) of 178 bits [15]. This is about 2.4 bit/second speed and that is about 2 times faster than the projected speed of our KLJN communicator for the same distance; note, the speed of KLJN communicator scales with the reciprocal of the range.

The speed of the present KLJN communicator is about 10 times slower than the theoretical limit for a single line (estimated in [1]). The speed can be increased and the theoretical limit can be approached by proper filtering, statistical tools, etc, to be described elsewhere. The only obvious way of significantly increasing the speed beyond the single line speed limit was already described in [1]. It is to use silicon chip technology with a large number of parallel communicators and multi-wire cables [1] (such as screened phone cables with 500 wires which are generally available). Such a solution can increase the speed by orders of magnitude however, due to cable costs, their practical use is very limited for very large distances, such as the case of 2000 km.

## 5. Conclusion

The realized KLJN communicator shows excellent performance through the model-line with communication ranges ten times beyond the range of known direct quantum communication channels with the longest range. The security claims of quantum communicators are usually theoretical, out of standard security measures provided by design and privacy application, because of the expensive nature of relevant breaking tests. However, in our practical KLJN system we have been able to test security at a much deeper level by carrying out security tests with various ways of breaking attempts.

The study indicates that the most important type of security leak, which may need most of resources (wire costs) to control, is the Bergou-Scheuer-Yariv type wire resistance effect [2,6,7]; and the rest of the proposed practical security compromises can be neglected, such as the transient/wave arguments of Scheuer-Yariv [6,7], the noise strength argument of Hao [8] and the resistor inaccuracy argument of Kish [9].

However, it is important to note that the information leak in the KLJN communicator is less harmful than the information leak in quantum communicators, namely, the eavesdropper does not know which KLJN bit is actually eavesdropped successfully. (In quantum communication the eavesdropper has a very good estimation about the location of the eavesdropped bits.)

The results indicate unrivalled fidelity and security levels among existing physical secure communicators and, if it is necessary and resources are available, there are straightforward ways to further improve security, fidelity and range, such as proper choice of resistors, thicker cable, enhanced statistical tools for bit decision, etc.

There is only one known practical situation of secure communication where only quantum solutions are favorable: the situation when wire communication is impossible but communication with photons can be done such as Zeilinger and coworkers' spatial quantum communication experiments mentioned above [15].

## Acknowledgements

Comments by Michael Weissman and Peter Makra are appreciated. Zoltan Gingl is grateful for the Bolyai Fellowship of Hungarian Academy of Sciences. The travel of LBK to the University of Szeged for the startup phase of the experiments was covered by the Swedish STINT foundation and the cost of staying (10-15 December, 2006) was partially covered by the European Union's SANES grant. The costs of the KLJN system design were partially covered by the TAMU Information Technology Task Force (TITF, grant 2002). Part of these results is also being reported in a plenary talk by the same authors at the Fluctuation and Noise Symposium (Florence, 2007) [14].